# REAL-TIME AUDIO TRANSLATION MODULE BETWEEN IAX AND RSW

Hadeel Saleh Haj Aliwi and Putra Sumari

School of Computer Sciences, Universiti Sains Malaysia, Penang, Malaysia

## ABSTRACT

*At the last few years, multimedia communication has been developed and improved rapidly in order to enable users to communicate between each other over the internet. Generally, multimedia communication consists of audio and video communication. However, this research concentrates on audio conferencing only. The audio translation between protocols is a very critical issue, because it solves the communication problems between any two protocols. So, it enables people around the world to talk with each other even they use different protocols. In this research, a real time audio translation module between two protocols has been done. These two protocols are: InterAsterisk eXchange Protocol (IAX) and Real-Time Switching Control Protocol (RSW), which they are widely used to provide two ways audio transfer feature. The solution here is to provide interworking between the two protocols which they have different media transports, audio codec's, header formats and different transport protocols for the audio transmission. This translation will help bridging the gap between the two protocols by providing interworking capability between the two audio streams of IAX and RSW. Some related works have been done to provide translation between IAX and RSW control signalling messages. But, this research paper concentrates on the translation that depends on the media transfer. The proposed translation module was tested and evaluated in different scenarios in order to examine its performance. The obtained results showed that the Real-Time Audio Translation Module produces lower rates of packet delay and jitter than the acceptance values for each of the mentioned performance metrics.*

## KEYWORDS

*Multimedia, VoIP, Interworking, InterAsterisk eXchange Protocol (IAX), Real Time Switching Control Criteria (RSW).*

## 1. INTRODUCTION

Multimedia communication has been developed and improved to be the basic and essential service in order to satisfy the needs of the Internet users [12, 18, 19, 20]. It has appeared to be more and more applicable in distributed environments [10, 21]. There are various protocols that control and signal the calls of the Internet telephone [1]; such protocols existed in Internet Protocol (IP) telecommunication. For data and signaling, two major protocols are considered in the field of the multimedia conferencing such as RSW and IAX protocols. Video functionality, similar quality of service and competitive voice are provided by these two protocols. The two protocols have been widely exploited and utilized into many different methods [15]. This work proposes a translation module between audio streams of IAX protocol and RSW control protocol. Generally, the procedure to a translator between the both protocols is starting from the first network to perform transferring the data into a first protocol. And then the second network to perform transferring the data into a second protocol. Finally, the translation server (conference gateway) to maintain translation information for a protocol translation, this translation occurs between the first and the second protocols [15].

IAX and RSW audio transfer protocols have many differences in handling and exchanging voice packets between each other. In addition, different transport protocols and techniques are used by each of these protocols throughout audio exchange. The packet format in IAX protocol differs from the corresponding packet in RSW control protocol. When the data is attempted to be exchanged among each other by the two protocols, the data could not be understood by each of the two protocols. To overcome the translation problem between the audio streams of the two protocols (IAX and RSW), we proposed an audio translation module, which will be able to act as a translator between IAX and RSW audio streams. This translation module will help bridging the gap between IAX and RSW by providing interworking capability between the two audio streams. There are two sides accompanied with the translation between the two protocols: the control signaling and audio stream translation. The control signaling translation between two protocols, more specifically between RSW control protocol and IAX protocol has already done [6,7]. And the audio stream translation, which is the proposed work in this paper. The performed audio stream translation is true while this research concentrates on the two protocols (RSW control protocol and IAX protocol).

## 2. BRIEF DESCRIPTION OF IAX AND RSW PROTOCOLS

This section will briefly describe the both protocols; Real-time Switching (RSW) Control Protocol and InterAsterisk eXchange (IAX) Protocol.

### 2.1. RSW Control Protocol

Real-time Switching (RSW) control criteria is a control protocol used to handle a multipoint-to-multipoint multimedia conferencing sessions. RSW control protocol was developed in 1993 as a control mechanism for conferencing by the Network Research Group in school of computer sciences, University Sciences Malaysia (USM) [9]. RSW protocol doesn't have its own header to carry the data so, Real-Time Protocol (RTP) protocol [3] is used to carry audio and video data through multimedia conferencing. User Datagram Protocol (UDP) is a transport protocol [2] that is also used by RSW to transfer audio and video data. The RSW control criteria is involved in decreasing bandwidth when many clients using the MCS system. RSW makes a list of priority for the participants to avoid confusion when many participants are trying to speak up during conference [6,13]. There are several advantages for the RSW control criteria [9] such as Equal Privileges, First Come First Serve, First come first serve with time-out, Organizer Main Site and Restricted Active site. The RSW audio packet format consists of four parts as shown in figure 1, which are IP header [2], UDP header, RTP header, and and RSW payload (varies based on the codec used) which is the body of the audio message.

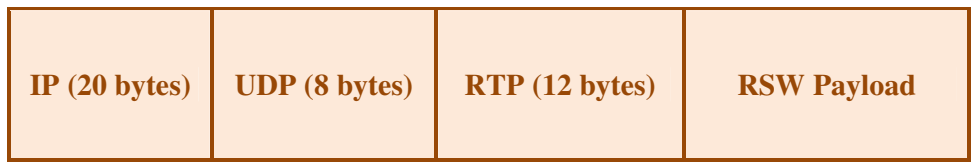

Figure 1.  RSW Audio Packet Format.

### 2.2. InterAsterisk eXchange Protocol

In (2004) Mark Spencer [5] has created the Inter-Asterisk eXchange (IAX) protocol for asterisk that performs VoIP signalling. Streaming media is managed, controlled and transmitted through the Internet Protocol (IP) networks based on this protocol. Any type of streaming media could be used by this protocol. However, IP voice calls are basically being controlled by IAX protocol [14]. Furthermore, this protocol can be called as a peer to peer (P2P) protocol that performs two types of connections which are Voice over IP (VoIP) connections through the servers and Client-Server communication. IAX is currently changed to IAX2 which is the second version of the IAX protocol. The IAX2 has deprecated the original IAX protocol [5]. Call signalling and multimedia transport functions are supported by the IAX protocol. In the same session and by using IAX, Voice streams (multimedia and signalling) are conveyed. Furthermore, IAX supports the trunk connections concept for numerous calls. The bandwidth usage is reduced when this concept is being used because all the protocol overhead is shared for all the calls between two IAX nodes. Over a single link, IAX provides multiplexing channels [11]. IAX is a simple protocol in such a way Network Address Translation (NAT) traversal complications are avoided by it [8]. Unlike RSW, IAX has its own frames (full and mini frames) to carry control, audio, video messages and signals. IAX full frames are responsible to carry the signals and the control messages, while Mini frame are the only type of frames which hold the data during the media transmission session between two endpoints A and B. Each audio/video flow is of IAX Mini Frames (M frames) contains 4 bytes header (source call number (2 bytes) and timestamp (2 bytes)). UDP transport protocol is used by IAX to transfer audio and video data [4]. The IAX audio packet format consists of four parts, which are the IP header, the UDP header, the IAX header [5] and the IAX payload (varies based on the codec used). Audio packet format of the IAX protocol is shown in figure 2.

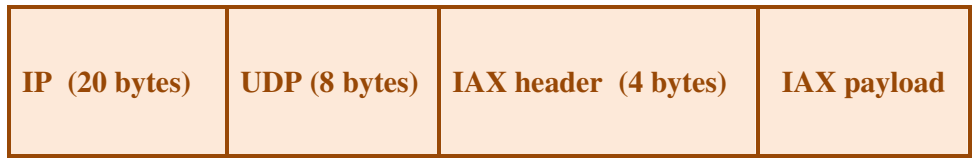

Figure 2. IAX Audio Packet Format.

## 3. CODEC'S USED IN IAX-RSW TRANSLATION MODULE

In order to transmit the packets from the sender to the receiver, an audio codec should be used. In this paper, GSM codec is proposed in order to convert the analog audio to digital form [17]. Then, the digital audio will be compressed to decrease the consumption of network bandwidth needed to pass on the speech to the receiver. GSM audio codec has the bandwidth of (13.2 kb/s). However, while sending the codec data across an IP network, it is going to increase the used bandwidth. While dealing with voice, it is better not to introduce too much latency. For example, sending voice frames every 20 ms, which means every 50 frames have to be sent in one second. Dividing 13.2 kb/s by 50 will give 33 bytes. As a result, 33 bytes of voice data will be sent per one frame. In addition, IP header (20 bytes a packet), UDP header (8 bytes a packet) adding the RTP protocol header (12 bytes a packet). This codec will be used not only for the IAX protocol, but also for the RSW control protocol. In this paper, the two audio streams protocols will use the same codec to transmit and receive the audio packets between the clients.

## 4. IAX-RSW TRANSLATION METHOD

This section will explain the two audio streams translation methods, which are RSW to IAX and IAX to RSW. The proposed translation module consists of three components: IAX network, RSW network, and the translation server or it can be called conference gateway. The proposed conference gateway includes two buffers; IAX to RSW buffer and RSW to IAX buffer. The main task of these buffers is converting the packet format of the source protocol to be identical to the format of the destination protocol to be clearly received.

- IAX to RSW buffer:

1. Collecting and storing the packets received from IAX client.

2. Extracting the mini header from IAX packets.

3. Inserting the RTP header (on behalf of the mini header) to IAX packets in order to be identical to RSW packet format.

4. Preparing the converted packets to be sent to the RSW client.

- RSW to IAX buffer:

1. Collecting and storing the packets received from RSW client.

2. Extracting the RTP header from RSW packets.

3. Inserting the mini header (on behalf of RTP header) to RSW packets in order to be identical to IAX packet format.

4. Preparing the converted packets to be sent to the IAX client.

## 5. SIMULATION ENVIRONMENT

Generally, there are no objections to have the flexibility to choose any simulation program to simulate the proposed translation module as long as it meets the requirements to simulate such a service. Furthermore, there are no restrictions to use particular hardware as a base to simulate the translation module. GloMoSim (Global Mobile system Simulator) was used to simulate the translation module [23]. GloMoSim is preferred to be chosen because it is extensible and composable [22]. The operating system that was used as an environment to run GloMoSim is Windows Vista. The hardware platform that was used to simulate the proposed translation module is Intel Core Duo 2.00 GHz.

The simulation model consists of three main elements, which are: IAX network, RSW network, and the conference gateway gateways with the internet connection. In the proposed simulation model, voice packets are collected in the gateway's buffer. The translation processes are done in the gateway. This simulation model has many parameters in order to understand the simulation processing, such as number of clients, buffer size, audio codec...etc.

## 6. PERFORMANCE METRICS TESTS

The real-time audio streams of IAX-RSW translation module will be evaluated by measuring and testing by two performance metrics, which are: Packet Delay and Jitter. In this scenarios, the packet delay and jitter tests of IAX-RSW translation module (one-to-one) were performed by taking one RSW client and one IAX client. Both ways (RSW-to-IAX and IAX-to-RSW) were tested using variable packet sequence numbers ranging from 10 to 100 packets.

## 6.1. Packet Delay Test

Packet Delay is defined as "The measure of time that it takes the talker's voice to reach the listener's ear" [24]. In the real-time applications, voice delay is considered to be a critical issue, such as audio conferencing systems [25].

The end-to-end acceptable delay is 150 milli secondsPacket delay examines the quality of VoIP service; as packet delay increased the QoS of VoIP calls decreased.

Packet delay is ranged between 1 and 15 ms in RSW-to-IAX transmission, while packet delay is ranged between 1 and 12 ms in IAX-to-RSW transmission.

As IAX protocol has trunking property, this makes the quality of services for this protocol are better than the other session protocols. Figure 3 shows that the IAX-RSW translation module result values for packet delay are less than the accepted packet delay value which is 150 ms.

## 6.2. Jitter Test

The transmitted packets from the source to the destination may be reached with different delays that rely on the queue in the crowded networks, several paths to avoid congestion, etc. This case cause dissimilar time variations.

The different time variations of received packets on the destination are known as Jitter. Jitter can bother and influence the quality of audio. Jitter should be less than 30 milliseconds [26].

Jitter examines the quality of VoIP service; as jitter increased the QoS of VoIP calls decreased.

According to figure 4, jitter is ranged between 2 and 4.3 ms in RSW-to-IAX transmission, while it is ranged between 2 and 3.9 ms in IAX-to-RSW transmission.

We can notice that the jitter results of both ways are almost convergent.

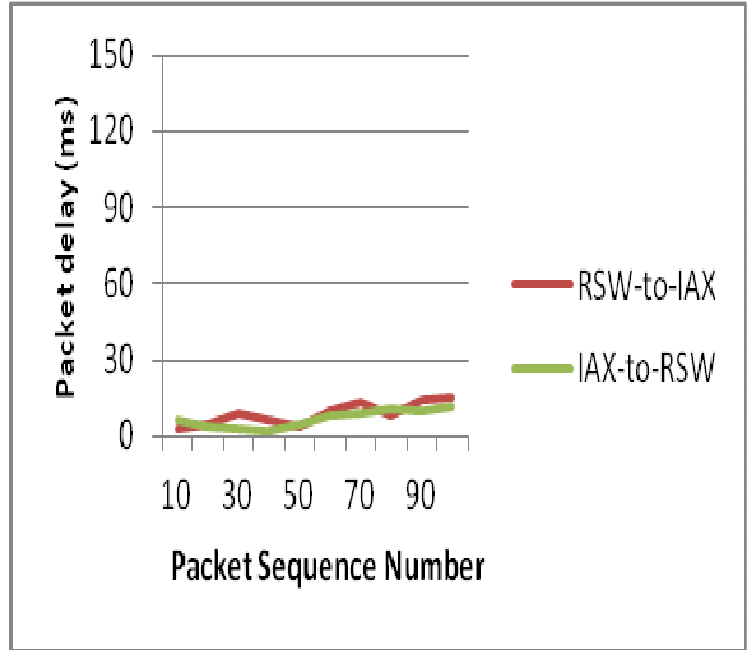

Figure 4.  One-to-One IAX-RSW Packet Delay Results.

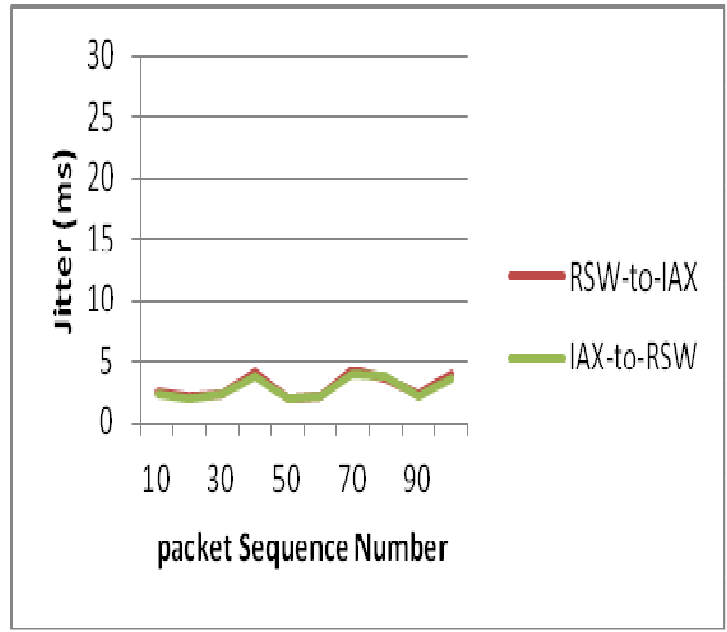

Figure 5.  One-to-One IAX-RSW Jitter Results.

# 7. SUMMARY

For guaranteeing a seamless end to end connectivity for IAX-RSW translation, we have proposed solution to these interworking problems by doing an audio translation module. Using the proposed Translation Module will enable VoIP protocols uses different sizes and formats of data chunks to exchange audio streams. The translation module can be a base for any two different protocols or more. The proposed method came as a translation audio streams module that can be used when provide IAX-RSW full interworking. It will be a valuable research if the translation module that depends on media transfer part (which is this performed work) combined with translation module that depends on the signalling part, so we can get completed results. The proposed translation module was tested and evaluated in different scenarios in order to examine its performance. The obtained results showed that the Real-Time Translation Module produces lower rates of packet delay and jitter than the acceptance values for each of the mentioned performance metrics.

**Authors**

**Hadeel Saleh Haj Aliwi** has obtained her Bachelor degree in Computer Engineering from Ittihad Private University, Syria in 2007-2008 and Master degree in Computer Science from Universiti Sains Malaysia, Penang, Malaysia in 2011. Currently, she is a PhD candidate at the School of Computer Science, Universiti Sains Malaysia. Her main research area interests are in includes Multimedia Networking, Voice over Internet protocols (VoIP), and Real-time Interworking between Heterogeneous signalling protocols, and Instant Messaging protocols. 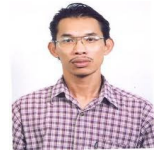

**Putra Sumari** obtained his MSc and PhD in 1997 and 2000 from Liverpool University, England. Currently, he is Associate Professor and a lecturer at the School of Computer Science, USM. He is the head of the Multimedia Computing Research Group, CS, USM. Member of ACM and IEEE, Program Committee and reviewer of several International Conference on Information and Communication Technology (ICT), Committee of Malaysian ISO Standard Working Group on Software Engineering Practice, Chairman of Industrial Training Program, School of Computer Science, USM, Advisor of Master in Multimedia Education Program, UPSI, Perak. 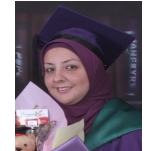